\documentstyle[12pt,psfig]{article}
\def\s{\star}
\begin{document}
\begin{titlepage}
\title{A clue to the mechanism of $\Lambda K^+$ production
in $pp$-reactions\thanks{Supported in part by the 
Forschungszentrum J\"ulich, and the 
Australian Research Council}}
\author{
A. Sibirtsev$^1$~\thanks{sibirt@theorie.physik.uni-giessen.de} ,
K. Tsushima$^2$~\thanks{ktsushim@physics.adelaide.edu.au} ,
A. W. Thomas$^2$~\thanks{athomas@physics.adelaide.edu.au} \\
{ $^1$\small Institut f\"ur Theoretische Physik, Universit\"at Giessen} \\
{\small D-35392 Giessen, Germany}  \\
{\small $^2$Department of Physics and Mathematical Physics } \\
{\small and Special Research Center for the Subatomic 
Structure of Matter} \\
{\small University of Adelaide, SA 5005, Australia} }
\date{}
\maketitle
\vspace{-10cm}
\hfill UGI-97-24, ADP-97-43/T271
\vspace{10cm}
\begin{abstract}
We analyse the $p p \to p \Lambda K^+$ total cross section data which were
recently measured by the COSY-11 collaboration at an energy  
2~MeV above the reaction threshold.
Our analysis suggests that the measurement of the invariant 
mass spectrum for the $\Lambda K^+$ at energies around 
100 MeV above the threshold can provide a new constraint 
on the theoretical calculations for this reaction. In particular, 
the measurement can give a clue as to whether the reaction 
is dominated by resonance excitation or not.   
Thus, it should contribute to further understanding of the  
strangeness production mechanism.
We show the invariant mass spectra for the $\Lambda K^+$ 
system in these optimal kinematic conditions 
calculated by several approaches, and demonstrate that our
suggestion is experimentally feasible.
\end{abstract}
\end{titlepage}
%

Recently the COSY-11 Collaboration~\cite{Balewski}
measured the total cross section for the reaction, 
$pp\to p \Lambda K^+$, at energy  
2~MeV above the reaction threshold. The data were compared 
with the {\it theoretical models}  which are presently 
available for this reaction~\cite{Sibirtsev,Li,Tsushima1,Wilkin}.
Among them, both the predictions of Sibirtsev~\cite{Sibirtsev} and
Li and Ko~\cite{Li} could reproduce the experimental data at this 
energy well only with the inclusion of pion and kaon exchanges, 
but without the inclusion of any resonances which are observed to decay to 
the $\Lambda K^+$ channel. 
Moreover, it was found~\cite{Sibirtsev,Li} that the dominant 
contribution to the $p p \to p \Lambda K^+$ reaction comes
from the $K$-exchange mechanism.
Alternatively, the study of F\"aldt and Wilkin~\cite{Wilkin} which 
was made analogous to the $p p \to p p \eta$ reaction 
with a nonrelativistic treatment including the final state interaction,  
suggests that the total cross sections for the 
$p p \to p \Lambda K^+$ reactions at energies only slightly above  
threshold can possibly be explained by one pion-exchange followed by 
$N^\s(1650)$ resonance excitation. Furthermore, they claim that 
these two ingredients alone are almost 
enough to describe the data. One of their motivations was   
to study whether the same type of meson exchange model could be capable
of explaining simultaneously the near-threshold
$\eta$ and $K$ production data -- one pion exchange followed by 
$N^\s(1650)$ resonance excitation for the present case.

\begin{figure}[hbt]
\psfig{figure=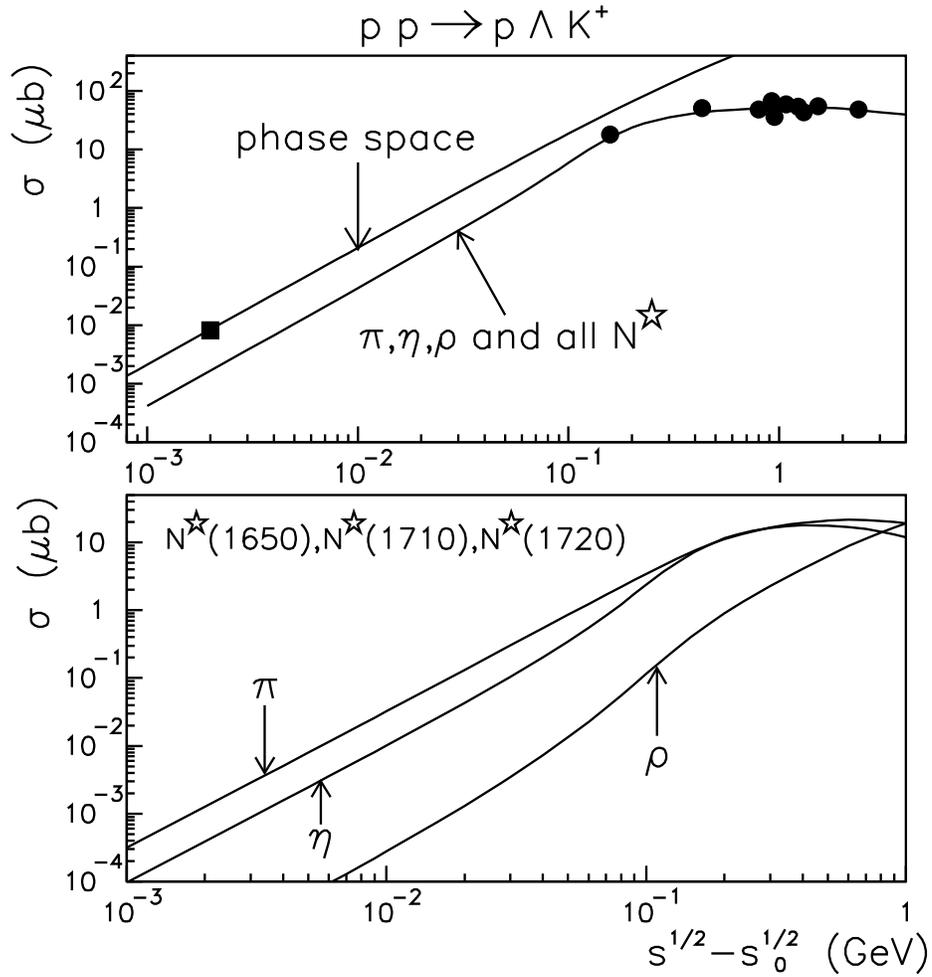,width=14cm}
\caption{\label{kt23p} Energy dependence of the total cross section
for the $p p \to p \Lambda  K^+$ reaction. The square indicates 
the experimental result of Ref.~\protect\cite{Balewski}, 
while the circles are 
data from Ref.~\protect\cite{Landolt}. In the upper part of figure, the 
solid lines indicate the results of the resonance model, 
and the phase space considerations using the 
constant matrix element normalized so as to fit the data.
The lower part in figure illustrates the 
separate contributions from $\pi$, $\rho$
and $\eta$-meson exchanges.}
\end{figure}

In Ref.~\cite{Tsushima1}, within the
resonance model, we performed the most detailed study of kaon
production in proton-proton collisions with a full relativistic 
treatment. This included the mechanism 
adopted by F\"aldt and Wilkin~\cite{Wilkin} as just one of the 
processes in evaluating the total cross sections.
However, it has turned out that 
our calculation underestimates the data~\cite{Balewski} 
at energy 2~MeV above the threshold -- as 
shown in Fig.~\ref{kt23p}. It underestimates the experimental
data from COSY by about a factor of 2, although it reproduces 
the existing data 
fairly well at energies more than 100 MeV above the threshold.
One of the reasons for this underestimation is that 
our calculation~\cite{Tsushima1} involves an incoherent 
sum of the contributions from each resonance, 
and thus would not usually be expected to reproduce the data well at 
energies very slightly above the threshold.

In Fig.~\ref{kt23p} we also show the energy dependence of 
the total cross sections calculated using  
a {\it constant} matrix element, adequately normalized 
so as to fit the data (or adequately scaled 
phase space distrubutions multiplied by a constant factor) for  
the reaction  -- denoted by $phase$ $space$.
It implies that the experimental data at low energies 
can be described solely by phase space with a constant 
matrix element normalized to fit the data, 
because a similar energy dependence of the production cross 
section is also obtained by both the $K$-meson exchange model of 
Refs.~\cite{Sibirtsev,Li} and the resonance model~\cite{Tsushima1}, 
except for the absolute values. Thus, the total cross section 
data for this reaction at very low energies might not be sufficient 
to constrain the theoretical calculations even at 
energies slightly above the threshold -- that is, to decide whether
the reaction is dominated by kaon exchange without resonance 
excitation~\cite{Sibirtsev,Li}, or by meson exchange  
followed by resonance excitation~\cite{Tsushima1,Wilkin}.  
Furthermore, it is difficult to decide 
what kind of meson exchange mechanism is dominant for the reaction as 
was questioned in Ref.~\cite{Wilkin}.

In the lower part of Fig.~\ref{kt23p}, we show the 
separate contribution from $\pi$, $\eta$ and $\rho$-meson 
exchanges to the energy dependence of the total cross sections.
In the calculation, we included the 
$N^\s(1650)$, $N^\s(1710)$ and $N^\s(1720)$
resonances. Their properties are summarized in Table~\ref{resonance}.
Most of the coupling constants, cut-off parameters and form factors 
have been fixed by the previous studies for the  
$\pi N \to Y K$ $(Y = \Sigma, \Lambda)$ reactions~\cite{Tsushima2,Tsushima4}.
Together with the quantities newly appearing in 
the $p p \to p \Lambda K^+$ reaction~\cite{Tsushima1}, 
we summarize the parameters of our model used for this study 
in Table~\ref{cconst}.

Here, we should comment that the main features of our results 
are very similar to those of F\"aldt and Wilkin~\cite{Wilkin}.
In particular, the $\pi$-meson exchange is dominant at energies close to 
the reaction threshold. Moreover, the $N^\s(1650)$ resonance is strongly 
coupled to the pion, and thus gives the main contribution for the 
reaction, which can also be understood from Table~\ref{resonance}.
Furthermore, it should be emphasized that we are able 
to reproduce the data 
of COSY-11~\cite{Balewski} if we adopt the upper limits for
the branching ratios for the $N^\s(1650)\to \pi N$ and 
$N^\s(1650) \to \Lambda K^+$ channels in Table~\ref{resonance}.
Indeed, the result was increased by a factor of about 1.8    
when we adopted the upper limits in the calculation.
Moreover, if we vary these branching ratios as 
free parameters, as was done 
by F\"{a}ldt and Wilkin~\cite{Wilkin}, 
we will be certainly able to reproduce the experimental 
data from COSY~\cite{Balewski}.

However, such approaches cause two problems.
First, with the same set of parameters, one should reproduce 
simultaneously the experimental data for the $\pi N \to \Lambda K$
reaction. Indeed, the approach of the resonance model 
made previously~\cite{Tsushima1} and again here uses the same parameter set 
consistently with that used for the study of the 
$\pi N \to \Lambda K$ reactions as well as the $\pi N \to \Sigma K$ 
reactions~\cite{Tsushima2,Tsushima4}.
Thus, our study for this reaction does not have much 
freedom to vary the parameters, nor to introduce new parameters. 
Second, the model should reproduce not only the recent 
data~\cite{Balewski}, but should also explain 
the data available at energies larger than 100~MeV 
above the threshold~\cite{Landolt}.

Keeping in mind the discussions made above, as well as 
the theoretical and experimental uncertainties  
for the $N^\s$ resonances, we may conclude that
{\it in principle} both the resonance model and  
the model of F\"{a}ldt and Wilkin~\cite{Wilkin} can 
reproduce the data for the $p p \to p \Lambda K^+$ at low energies, 
although this was not our motivation, nor the intention 
of the calculation~\cite{Tsushima1}.

Here, we point out further that a more definite conclusion can be drawn 
from the measurement of the $\Lambda K^+$ invariant mass spectrum.
This might also clarify whether the other resonances, $N^\s(1710)$
and $N^\s(1720)$, may play a role in the reaction as well as the
$N^\s(1650)$ resonance at relatively low energies near the threshold. 

\begin{figure}[hbt]
\psfig{figure=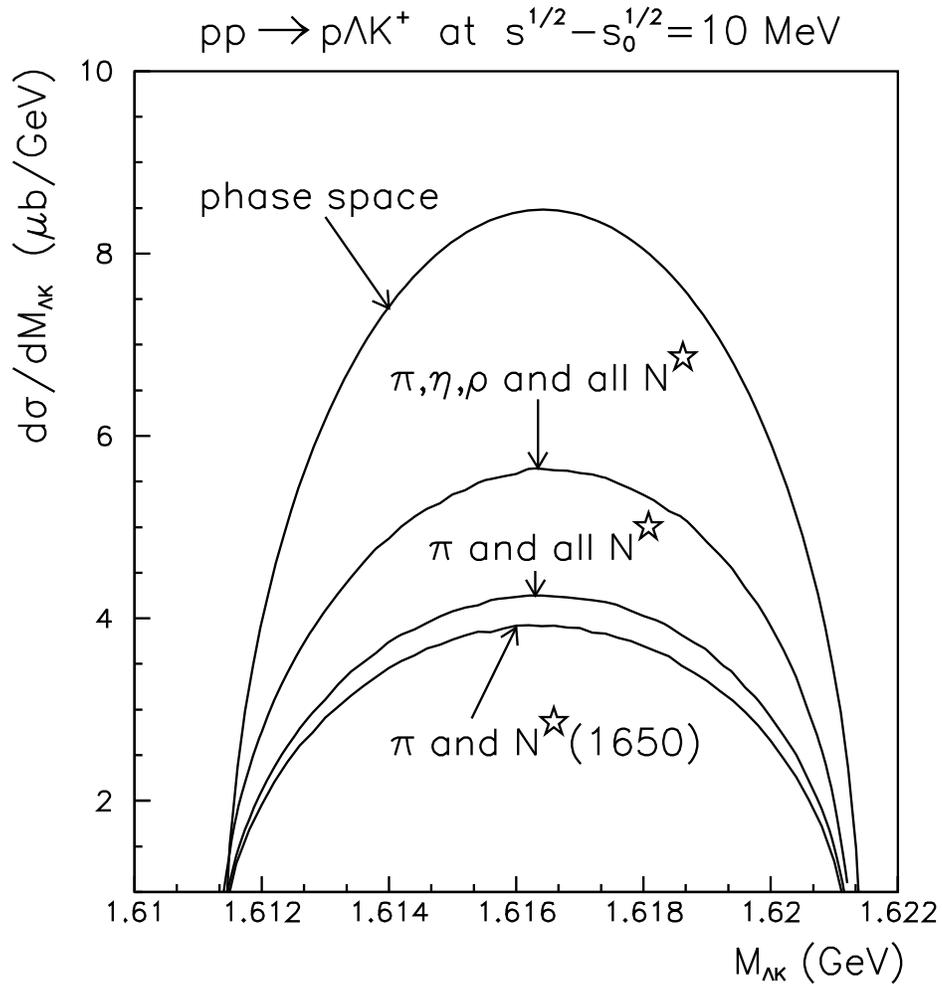,width=14cm}
\caption{\label{kt22p} The $\Lambda K^+$ invariant mass spectrum 
for the $p p \to p \Lambda  K^+$ reaction.
Results are shown for the resonance model, and a simple phase space 
consideration with an arbitrary normalization. 
Here the reaction energy is 10~MeV above the threshold.}
\end{figure}

When the $\Lambda K^+ $ pair is produced through 
$N^\s$ resonance excitations, the 
invariant mass spectrum of the $\Lambda K^+$ pair is 
expected to be influenced by the resonances\footnote{Note that expected  
mass spectrum is not a simple Breit-Wigner 
distribution~\cite{Byckling}.}. 
In Fig.~\ref{kt22p} we first show the spectrum of 
the $\Lambda K^+$ invariant mass
calculated with the resonance model at a excess enegy, 
$\epsilon = \sqrt{s}-\sqrt{s_0}=10$~MeV, 
where $\sqrt{s_0}$ is the reaction 
threshold, namely $\sqrt{s_0} = m_N + m_\Lambda + m_K = 2.5503$ GeV 
with $m_N$, $m_\Lambda $  and 
$m_K$ being the corresponding masses 
of the particles appearing in the final state. 
We show the spectrum calculated for the following three cases:  
1) the total contribution from $\pi$, $\eta$ and $\rho$-meson 
exchanges with the inclusion of all the resonances, 
$N^\s(1650)$, $N^\s(1710)$ and $N^\s(1720)$,  
2) the separate contribution from pion exchange alone with 
the inclusion of all the $N^\s$ resonances, and 3) pion exchange with 
the $N^\s(1650)$ resonance alone.
For comparison we also illustrate the phase space distributions
with an arbitrary normalization. It is easily seen that
the shapes of all spectra for the three cases at this energy 
are similar to that of simple phase space.

\begin{figure}[hbt]
\psfig{figure=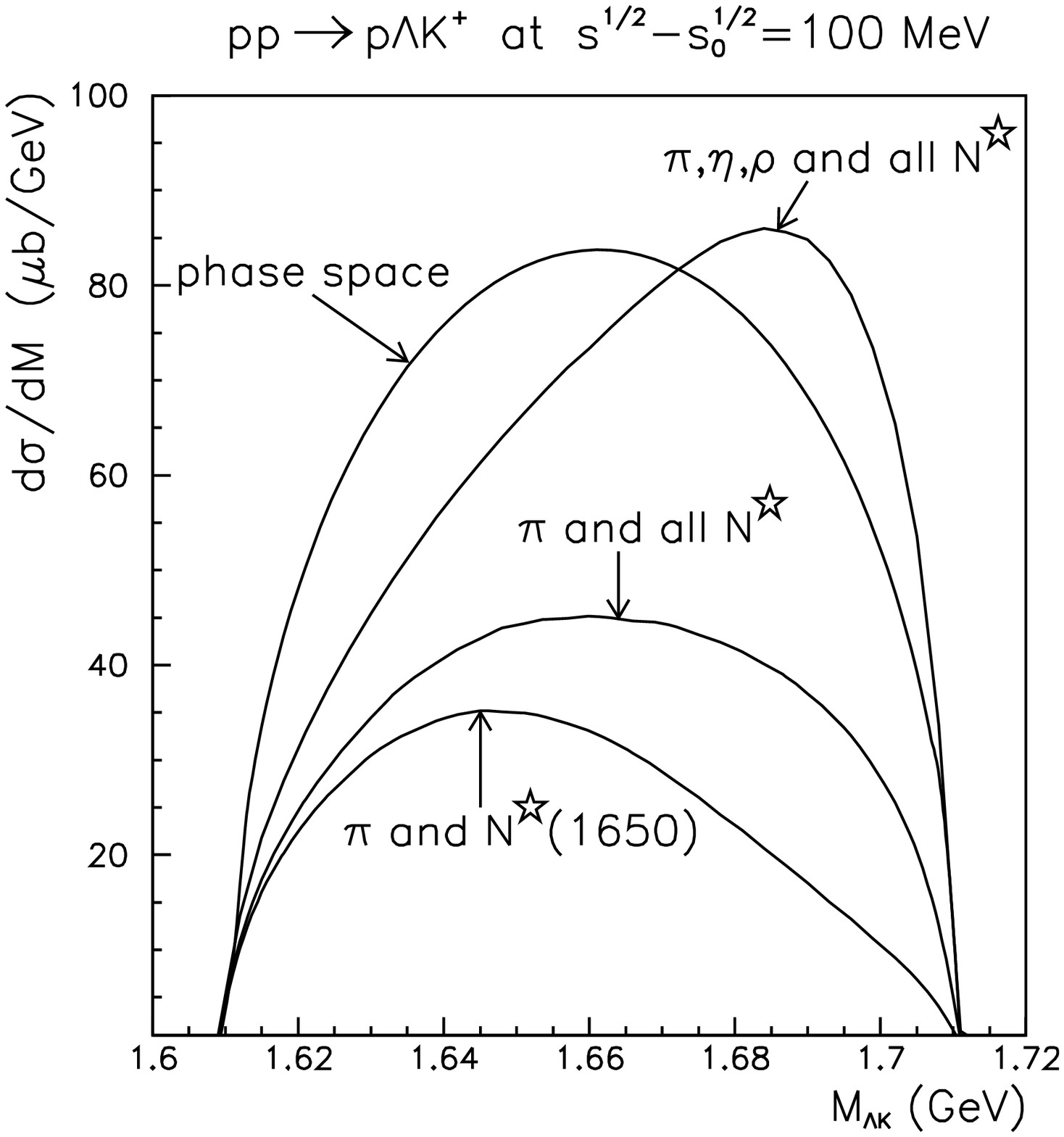,width=14cm}
\caption{\label{kt21p} Same as Fig.~\protect\ref{kt22p}, but 
the reaction energy is 100~MeV above the threshold.}
\end{figure}

The reason is that the width of each $N^\s$ resonance is larger than the
maximal total energy (when the proton in the final state has zero momentum) 
available for the $\Lambda K^+ $ pair, 
$\epsilon = 10$~MeV. Because the upper limit of the spectrum is 
given by, $\sqrt{s_0} + \epsilon - m_p = m_\Lambda + m_K + \epsilon$,  
the range of the invariant mass distribution (i.e. the
difference between the upper and lower limit for the spectrum)
is equal to the excess energy, $\epsilon$, above threshold. 
Thus, for a small value of the excess energy, $\epsilon$,  the structure
of the corresponding resonance -- in the present case the lowest 
baryonic resonance, $N^\s(1650)$ -- cannot affect the shape
of the $\Lambda K^+$ invariant mass spectrum noticeably, because 
the width is much larger than the excess energy.  
Obviously when the quantity, $ m_\Lambda + m_K + \epsilon$, 
is larger than the 
value, $m_{N^\s(1650)} + {\Gamma}_{N^\s(1650)}/2$, 
the mass spectrum will be certainly affected by this resonance -- where, 
$m_{N^\s(1650)}$ and ${\Gamma}_{N^\s(1650)}$ are respectively the
mass and width of the $N^\s(1650)$ resonance.
Then, if resonance excitation is the dominant mechanism 
for the $p p \to p \Lambda K^+$ reaction, we expect that the  
$\Lambda K^+$ invariant mass spectrum will be certainly affected
by the baryonic resonances under the condition, 
$\epsilon \simeq m_{N^\s(1650)} + {\Gamma}_{N^\s(1650)}/2 
- (m_\Lambda + m_K) \simeq 100$ MeV. (The $N^\s(1650)$ resonance 
will at first start to give a contribution as the energy increases.) 
In order to demonstrate that 
our argument is correct, 
we illustrate the $\Lambda K^+$ invariant mass spectrum
calculated at an excess energy of 100~MeV 
in Fig.~\ref{kt21p}, as an example.
As can be seen from Fig.~\ref{kt21p}, 
the contributions of $\pi$ and $\eta$-meson
exchange become important when all the resonances, 
$N^\s(1650)$, $N^\s(1710)$ and $N^\s(1720)$ are included together.
Furthermore, we emphasize that the result of the resonance model 
can be now distinguished from those of the simple phase space spectrum 
with an arbitrary normalization, and the pion exchange with the 
$N^\s(1650)$ resonance alone in our model.
Thus, the measurement of the $\Lambda K^+$ invariant mass spectrum 
may serve as a decisive test for the theoretical models and 
the mechanism for strangeness production.

It is of great interest that a measurement of assosiated
strangeness production in proton-proton collisions is
planned by the COSY-TOF Collaboration~\cite{TOF} in the near future.
It is also crucial, to test our suggestion, 
that the experiment can be performed
with a clear identification of the $\Lambda$ and $\Sigma$ 
hyperons in the invariant mass  
spectra -- as was emphasized by Laget~\cite{Laget}.\\ \\

\noindent 
Acknowledgement\\
A.S. would like to thank L.~Jarczyk, W.~Cassing, B. Kamys
and U.~Mosel for productive discussions.
This work was supported in part by the Forschungszentrum J\"{u}lich, 
and the Australian Research Council.

%
\newpage
\begin{table}
\caption{\label{resonance} Summary of the resonance properties used in the 
present study. The confidence levels of the resonances are,
$N^\s(1650)****, N^\s(1710)***$ and 
$N^\s(1720)****$~\protect\cite{particle}.}
\begin{center}
\begin{tabular}{|c|c|c|c|c|}
\hline
Resonance $(J^P)$ &Width (MeV) &Decay channel &Branching ratio 
&Adopted value \\
\hline
$N^\s(1650)\,(\frac{1}{2}^-)$ &150 &$N \pi$      &0.60 -- 0.80 &0.700 \\
                           &    &$N \eta$     &0.03 -- 0.10 &0.065 \\
                           &    &$\Lambda K$  &0.03 -- 0.11 &0.070 \\
\hline
$N^\s(1710)\,(\frac{1}{2}^+)$ &100 &$N \pi$      &0.10 -- 0.20 &0.150 \\
                           &    &$N \eta$     &0.20 -- 0.40 &0.300 \\
                           &    &$N \rho$     &0.05 -- 0.25 &0.150 \\
                           &    &$\Lambda K$  &0.05 -- 0.25 &0.150 \\
                           &    &$\Sigma K$   &0.02 -- 0.10 &0.060 \\
\hline
$N^\s(1720)\,(\frac{3}{2}^+)$ &150 &$N \pi$      &0.10 -- 0.20 &0.150 \\
                           &    &$N \eta$     &0.02 -- 0.06 &0.040 \\
                           &    &$N \rho$     &0.70 -- 0.85 &0.775 \\
                           &    &$\Lambda K$  &0.03 -- 0.10 &0.065 \\
\hline
\end{tabular}
\end{center}
\end{table}
%
\begin{table}
\caption{\label{cconst}
Coupling constants and cut-off parameters. 
$\kappa = f_{\rho N N}/g_{\rho N N} = 6.1$ 
for the $\rho N N$ tensor coupling is used.} 
\begin{center}
\begin{tabular}{|l|l|c|c|l|l|c|}
\hline
vertex & $g^2/4\pi$ & cut-off (MeV)\\
\hline 
 & & \\
$\pi N N$  & $14.4$ & $1050$ \\ 
$\pi N N(1650)$ & $1.12 \times 10^{-1}$ & $800$ \\
$\pi N N(1710)$ &$2.05 \times 10^{-1}$ &$800$ \\
$\pi N N(1720)$ &$4.13 \times 10^{-3}$ &$800$ \\
$\eta N N$ &$5.00$ &$2000$\\
$\eta N N(1650)$ &$3.37 \times 10^{-2}$ &$800$\\ 
$\eta N N(1710)$ &$2.31$ &$800$ \\
$\eta N N(1720)$ &$1.03 \times 10^{-1}$ &$800$\\
$\rho N N$  &$0.74$ &$920$ \\ 
$\rho N N(1710)$ &$3.61 \times 10^{+1}$ &$800$\\
$\rho N N(1720)$  &$1.43 \times 10^{+2}$ &$800$ \\ 
$K \Lambda N(1650)$ &$5.10 \times 10^{-2}$ &$800$\\
$K \Lambda N(1710)$ &$3.78$ &$800$ \\
$K \Lambda N(1720)$ &$3.12 \times 10^{-1}$ &$800$\\
 & & \\ 
\hline
\end{tabular}
\end{center}
\end{table}
%
\newpage
%
%

\begin{thebibliography}{99}
\bibitem{Balewski} 
        J. Balewski et al., Phys. Lett. B 388 (1996) 859.
\bibitem{Sibirtsev} 
        A. Sibirtsev, Phys. Lett. B 359 (1995) 29.
\bibitem{Li}
        G.Q. Li and C.M. Ko, Nucl. Phys. A 594 (1995) 439;\\
        See, G.Q. Li, C.-H. Lee, and G.E. Brown, nucl-th/9706057, 
        for comparion of their calculation and the experimental data.
\bibitem{Tsushima1}
        K. Tsushima, A. Sibirtsev, A.W. Thomas, Phys. Lett.
        B 390 (1997) 29.
\bibitem{Wilkin}
        G. F\"aldt and C. Wilkin, Z. Phys. A 357 (1997) 241.
\bibitem{Tsushima2} 
        K. Tsushima, S.W. Huang and A. Faessler, Phys. Lett.
        B 337 (1994) 245.
\bibitem{Tsushima3} 
        K. Tsushima, S.W. Huang and A. Faessler, J. Phys.
        G 21 (1995) 33.
\bibitem{Tsushima4} 
        K. Tsushima, S.W. Huang and A. Faessler, nucl-th/9602005, 
        Australian J. Phys. 50 (1997) 35.
\bibitem{Landolt} 
        Landolt-B{\"{o}}rnstein, New Series, 
        ed. H. Schopper, I/12 (1988). 
\bibitem{Byckling}
        E. Byckling, K. Kajantie,  Particle Kinematics, 
        John Wiley \& Sons, 1973.
\bibitem{TOF} 
        W. Eyrich et al., Annual Report 1996 J\"ul-3365 p.7.
\bibitem{Laget} 
        J.M. Laget, Phys. Lett. B 259 (1991) 24.
\bibitem{particle} 
        Particle Data Group, Phys. Rev. D 50 (1994).
\end{thebibliography}
\end{document}